%% file: main.tex
\documentclass[conference]{IEEEtran}
\IEEEoverridecommandlockouts
\usepackage{booktabs}
\usepackage{graphicx} %
\usepackage[acronym]{glossaries}
\usepackage{glossaries-extra}
\glsdisablehyper %
\setabbreviationstyle[acronym]{long-short}
\usepackage{xspace}
\usepackage{xcolor}
\usepackage{todonotes}
\usepackage{hyperref}
\hypersetup{breaklinks=true}
\usepackage{cleveref}
\usepackage{versions}
\usepackage[backend=biber,style=ieee]{biblatex}
\usepackage{listings}
\usepackage[frozencache=true,cachedir=minted-cache,newfloat=true]{minted}
\usepackage{textcomp}

\bibliography{refs-blime,more-refs}

\excludeversion{paper}
\includeversion{paper}

\input{macros}
\input{glossary}

\begin{document}

\title{BliMe Linter}

\author{\IEEEauthorblockN{Hossam ElAtali\IEEEauthorrefmark{1}, Xiaohe Duan\IEEEauthorrefmark{1}, Hans Liljestrand\IEEEauthorrefmark{2}, Meng Xu\IEEEauthorrefmark{1} and N. Asokan\IEEEauthorrefmark{1}}
\IEEEauthorblockA{\IEEEauthorrefmark{1}University of Waterloo\\
hossam.elatali, x34duan, meng.xu.cs@uwaterloo.ca\\
asokan@acm.org}
\IEEEauthorblockA{\IEEEauthorrefmark{2}Intel Labs\\
hans@liljestrand.dev}}

\maketitle
\thispagestyle{plain}
\pagestyle{plain}

\input{sections/0_abstract}
\glsresetall
\input{sections/1_intro}
\input{sections/2_background}
\input{sections/4_design}

\input{sections/5_eval}

\input{sections/6_discussion}

\input{sections/7_related}

\input{sections/8_conclusion}

\printbibliography

\end{document}

%% file: macros.tex
\newcommand{\design}{the BliMe linter\xspace}
\newcommand{\Design}{The BliMe linter\xspace}

%% file: glossary.tex
\newacronym{ABI}{ABI}{application binary interface}
\newacronym{API}{API}{application programming interface}
\newacronym{BliMe}{BliMe}{Blinded Memory}
\newacronym{BliMe-Ibex}{\textsf{BliMe-Ibex}}{Blinded Memory for Ibex}
\newacronym{BliMe-BOOM}{\textsf{BliMe-BOOM}}{Blinded Memory for BOOM}
\newacronym{CPU}{CPU}{central processing unit}
\newacronym{DIFT}{DIFT}{dynamic information flow tracking}
\newacronym{IFT}{IFT}{information flow tracking}
\newacronym{FHE}{FHE}{fully-homomorphic encryption}
\newacronym{FPGA}{FPGA}{field-programmable gate array}
\newacronym{IPC}{IPC}{inter-process communication}
\newacronym{IR}{IR}{intermediate representation}
\newacronym{ISA}{ISA}{instruction set architecture}
\newacronym{LUT}{LUT}{look-up table}
\newacronym{PIR}{PIR}{private information retrieval}
\newacronym{RAM}{RAM}{random access memory}
\newacronym{REE}{REE}{rich execution environment}
\newacronym{SoC}{SoC}{system-on-chip}
\newacronym{TA}{TA}{trusted application}
\newacronym{TCB}{TCB}{trusted code base}
\newacronym{TEE}{TEE}{trusted execution environment}
\newacronym{OS}{OS}{operating system}
\newacronym{DVFS}{DVFS}{dynamic voltage and frequency scaling}
\newacronym{OMP}{OMP}{oblivious memory partition}
\newacronym{OISA}{OISA}{oblivious instruction set architecture}
\newacronym{SEP}{SEP}{Secure Enclave Processor}
\newacronym{DRAM}{DRAM}{dynamic random-access memory}
\newacronym{HSM}{HSM}{hardware security module}
\newacronym{STT}{STT}{Speculative Taint Tracking}
\newacronym{PTE}{PTE}{page table entry}
\newacronym{CSP}{CSP}{cloud service provider}
\newacronym{SVFG}{SVFG}{sparse value-flow graph}
\newacronym{CFL}{CFL}{control-flow linearization}
\newacronym{DFL}{DFL}{data-flow linearization}
\newacronym{wllvm}{WLLVM}{whole-program LLVM}

%% file: sections/0_abstract.tex
\begin{abstract}

\begin{paper}
Outsourced computation presents a risk to the confidentiality of clients' sensitive data since they have to trust that the service providers will not mishandle this data. \gls{BliMe}~\cite{blime24} is a set of hardware extensions that addresses this problem by using hardware-based taint tracking to keep track of sensitive client data and enforce a security policy that prevents software from leaking this data, either directly or through side channels. Since programs can leak sensitive data through timing channels and memory access patterns when this data is used in control-flow or memory access instructions, \gls{BliMe} prohibits such unsafe operations and only allows \emph{constant-time code} to operate on sensitive data. The question is how a developer can confirm that their code will run correctly on BliMe. While a program can be manually checked to see if it is constant-time, this process is tedious and error-prone.

In this paper, we introduce \design{}, a set of compiler extensions built on top of SVF~\cite{SVF} 
that analyze LLVM bitcode to identify possible \gls{BliMe} violations%
.
We evaluate \design{} analytically and empirically and show that it is sound.
\end{paper}
\end{abstract}

%% file: sections/1_intro.tex
\section{Introduction}
Outsourced computing offers efficient and stable platforms and services for individuals and organizations to handle data processing tasks. Typically, in this setup, clients entrust service providers with their sensitive data, which is then processed and the results are returned. However, there is a potential security risk because the software underlying outsourced computing infrastructure can be malicious or vulnerable to external threats. Consequently, client data can be exposed through run-time attacks or more concealed leakage like side-channel attacks. Addressing this concern about the leakage of sensitive data is a significant challenge in outsourced computing.

Many approaches have been proposed to mitigate data leaks in outsourced computing~\cite{yu19, intelSGXExplained2016, dove, gentryFHE}. One state-of-the-art solution is \gls{BliMe}~\cite{blime24}, which is a set of hardware extensions that enforce a taint-tracking policy to ensure data confidentiality. The policy monitors data that depends on secrets and triggers a fault if an instruction may reveal secret-dependent data to external observers, whether through run-time attacks or side channels. BliMe ensures that decrypted data on the server-side is always tagged as tainted (including in registers, caches and memory) and is tracked by the hardware, thus preventing potential leaks from vulnerable or malicious server-side software. Unlike \gls{FHE}, BliMe allows the processor to conduct computations directly on decrypted data after importing it, avoiding the high performance overhead associated with operating on encrypted data.

While BliMe hardware provides efficient and secure outsourced computation, it has a stringent policy that aborts any secret-dependent branching or memory access instructions to prevent adversaries from inferring secrets by monitoring execution times. This requirement presents a usability challenge because most software is not designed to be constant-time or BliMe-compliant. In order for such software to run on BliMe, its executable needs to be adapted into a form that avoids secret-dependent control-flow branches and memory accesses%
. Since manually identifying potential data leaks is a complex and error-prone task, a compiler-based tool is needed to automatically and comprehensively identify potential violations in source code.

\begin{paper}
In this paper, we introduce \design{}, a set of compiler extensions that analyze LLVM bitcode to identify possible \gls{BliMe} violations. At the core of our constant-time code linter is a taint-tracking engine that propagates taint statically but with the same policy as enforced by the BliMe hardware at runtime. To detect information flows within the program, we build on a state-of-the-art static analysis tool, SVF~\cite{SVF}, that uses value flows to track relations in programs (\Cref{sec:SVF}). SVF alone, however, is not enough to detect all BliMe violations. We discuss the reasons for this, as well as how we improve the analysis to account for it, in \Cref{sec:taint-tracking}. Concretely, our contributions are as follows.
\begin{enumerate}
    \item \Design{}, a set of compiler extensions to identify potential \gls{BliMe} violations in LLVM bitcode (\Cref{sec:design}).\footnote{\Design{} will be open-sourced on publication.}
    \item An evaluation of \design{} on the \gls{OISA} benchmarks and TensorFlow Lite (\Cref{sec:eval}).
    \item Two case studies that demonstrate the effectiveness of \design{} in identifying the root cause of violations (\Cref{sec:case-studies}).
    \item A discussion of challenges and possible improvements for future work, including automatic transformations of non-compliant code (\Cref{sec:discussion}).
\end{enumerate}
\end{paper}

%% file: sections/2_background.tex
\section{Background}
\subsection{Side channels}
Side channels capture side effects of program execution which are not observed through the intended media. Examples of \gls{CPU} side channels include execution time, memory access patterns, microarchitectural state (e.g., caches), voltage and electromagnetic radiation~\cite{paccagnellaLordRingSide2021,kocher99dpa,emAttack1,emAttack2,lippPLATYPUSSoftwarebasedPower2021,kimThermalBleedPracticalThermal2022}. Side-channel leakage occurs when an adversary can deduce information about secret data by observing the system while it is being processed. One possible cause is vulnerabilities in the underlying hardware, e.g., Meltdown~\cite{Lipp2018meltdown}. Another is secret-data-dependent behavior by the program. For instance, if a program's execution time varies depending on a sensitive value used in a conditional branch instruction, an adversary could infer details about that value by monitoring the branch's completion time. Another scenario is when a sensitive value is used to access an array in memory, causing a change in the memory access pattern, which can be observed through shared caches or memory buses.

Performance optimizations common in modern \glspl{CPU} can further amplify side-channel leakage~\cite{Lipp2018meltdown,Kocher2018spectre,schwarzZombieLoadCrossprivilegeboundaryData2019,vanbulck2018foreshadow,ragabRageMachineClear2021,vanbulck2018foreshadow,gotzfriedCacheAttacksIntel2017a,brasserSoftwareGrandExposure2017,leeInferringFinegrainedControl2017,chenSgxPectreStealingIntel2019,ryanHardwarebackedHeistExtracting2019,zhangTruSenseInformationLeakage2018}. 
This behavior can occur transparently under-the-hood, making it difficult for developers to identify vulnerabilities in programs.

\subsection{Constant-time code}\label{sec:constant-time}
Constant-time programming is a programming paradigm widely used in cryptographic libraries to mitigate side-channel leakage. The idea is to prevent sensitive data from being used in ways that affect execution time, which can manifest in two ways: control flow and data flow. The techniques employed to prevent such leakage are called \gls{CFL} and \gls{DFL}, respectively.

For control flow, state-of-the-art solutions use a program counter security model (PC-security)~\cite{molnarProgramCounterSecurity2005}. PC-security states that the program counter cannot become sensitive-data-dependent at any point during the execution of the program. This effectively means that an adversary cannot use a trace of the program's execution to infer sensitive data values. PC-security prevents using sensitive data to select conditional branches or determine whether a fault occurs (e.g., division by zero). \Gls{CFL} attempts to provide PC-security by executing both paths of a sensitive conditional branch (a real path and a decoy path), while maintaining correct functionality of the program. \emph{Predicated execution}, a form of \gls{CFL}, does this by maintaining a predicate throughout execution that represents whether this path is a real or decoy path. Every sensitive operation is then masked with this predicate to ensure that it only changes the program state if it is on the real path. \emph{Transactional execution}, another form of \gls{CFL}, executes both paths as-is but attempts to buffer and then discard state updates from decoy paths.

PC-security alone, however, is not sufficient. Sensitive-data-dependent data flows can also affect execution time. For example, some \gls{CPU} implementations of floating point instructions have a variable number of cycles that depends on the operand values. If sensitive values are used as operands to such instructions, the execution time (in cycles) can leak the values to an adversary. Furthermore, memory accesses can have different latencies depending on whether the requested address is cached. A leak can therefore also occur if the addresses of memory accesses are sensitive. One simple \gls{DFL} technique for instructions with a variable number of cycles is software emulation; the unsafe \gls{CPU} instruction is replaced with a safe instruction sequence that performs the same functionality. For memory accesses, \gls{DFL} is more difficult. The access latency must be indistinguishable for the set of all possible addresses that could be accessed at this point in the program execution. For arrays, one way to achieve this is simply by fetching the entire array into the cache before each access. This ensures that each access, irrespective of the sensitive address, will always have the cache-hit latency. Another more general approach is to identify, for each memory instruction, the set of all possible memory addresses that can be accessed by the instruction, and simply access the entire set, masking away all values except the desired one.

\subsection{SVF}\label{sec:SVF}
In order to facilitate constant-time programming, compilers must determine which values are sensitive and how they affect or propagate to other values. One way to do this is by performing static program analysis, specifically value flow analysis. Value flow analysis resolves dependencies between variables in the program. SVF~\cite{SVF} is a state-of-the-art value-flow analysis tool. It splits the analysis into two steps that are performed iteratively: pointer analysis, and value-flow construction. Pointer analysis determines the set of locations each pointer can refer to. Value-flow construction takes the results of the pointer analysis and uses it to create a \gls{SVFG}, which represents the dependencies between values as a directed graph. If further precision is required, the \gls{SVFG} can be fed back to the pointer analysis step to improve its precision, resulting in further improvement to the \gls{SVFG}.

\subsection{BliMe}
BliMe aims to enable clients to securely transmit sensitive data to a server for processing, ensuring that the data remains confidential and is not leaked through direct means or side channels. It comprises a minimal set of hardware extensions and an attestation architecture. This setup allows a client to send conventionally encrypted data to a remote server, which the processor can decrypt and process without allowing the data or any derived information to be leaked. Results are only returned as ciphertext after encrypting them with the client's key.

In BliMe, the processor prevents client data from being exported from the system by enforcing a taint-tracking policy.
Secure import and export of data between clients and servers is provided by using a \gls{HSM} and an encryption engine, which ensure that decrypted client data on the server is always tainted. This allows BliMe to provide its security guarantees \emph{without having to make any trust assumptions about the server code}. As a result, BliMe ensures client data confidentiality even against run-time attacks and malicious server code.

BliMe raises a fault when there is any attempt by software to leak blinded data. As a result, programs must be ``BliMe-compliant'' to run without being halted by BliMe.

\subsubsection{BliMe-compliant vs. Constant-time}
BliMe enforces its policy on a per-instruction basis and does not have a high-level view of program semantics. Therefore, it enforces a stricter form of constant-timeness compared to that discussed in \Cref{sec:constant-time}%
. For example, traditional constant-time code allows branching on secret data as long as both branches are balanced and no timing difference can be detected between their executions. Another example is when software strides arrays with cache-line-sized step and a secret offset within each cache line (\Cref{sec:array-access}). In both cases, data confidentiality guarantees stem from the software and the developer must reason about the software's security to provide these guarantees. BliMe, on the other hand, does not rely on trust in software to provide its security guarantees, and therefore must take a conservative approach. As a result, both cases discussed above are prohibited by BliMe.

\subsubsection{BliMe vs. compiler taint tracking}\label{sec:blime-vs-compiler}
Taint tracking in BliMe is dynamic and fully precise. The hardware can track exactly how tainted values are used and how the taint propagates in registers and memory. Consequently, BliMe precisely knows the taint of a program in its current execution state. On the other hand, static taint analysis attempts to determine the taint for all possible execution states at a specific program location. This is an undecidable problem~\cite{riceClasses1953}, and therefore compilers must make approximations to keep the problem tractable. These approximations must be conservative, which can lead to an over-approximation of the final tainted state. We discuss how the analysis can be made more precise in \Cref{sec:overapprox}.

%% file: sections/4_design.tex
\section{Design \& Implementation}\label{sec:design}

The main requirement for \design{} is that the analysis must be sound but not necessarily complete. This means that the analysis %
may fail to mark certain BliMe-compliant code as safe, but code reported as safe by the linter must have no violations on the actual BliMe hardware.

\begin{figure}
    \centering
    \includegraphics[width=\linewidth,trim={2.5cm 4.5cm 3cm 4cm},clip]{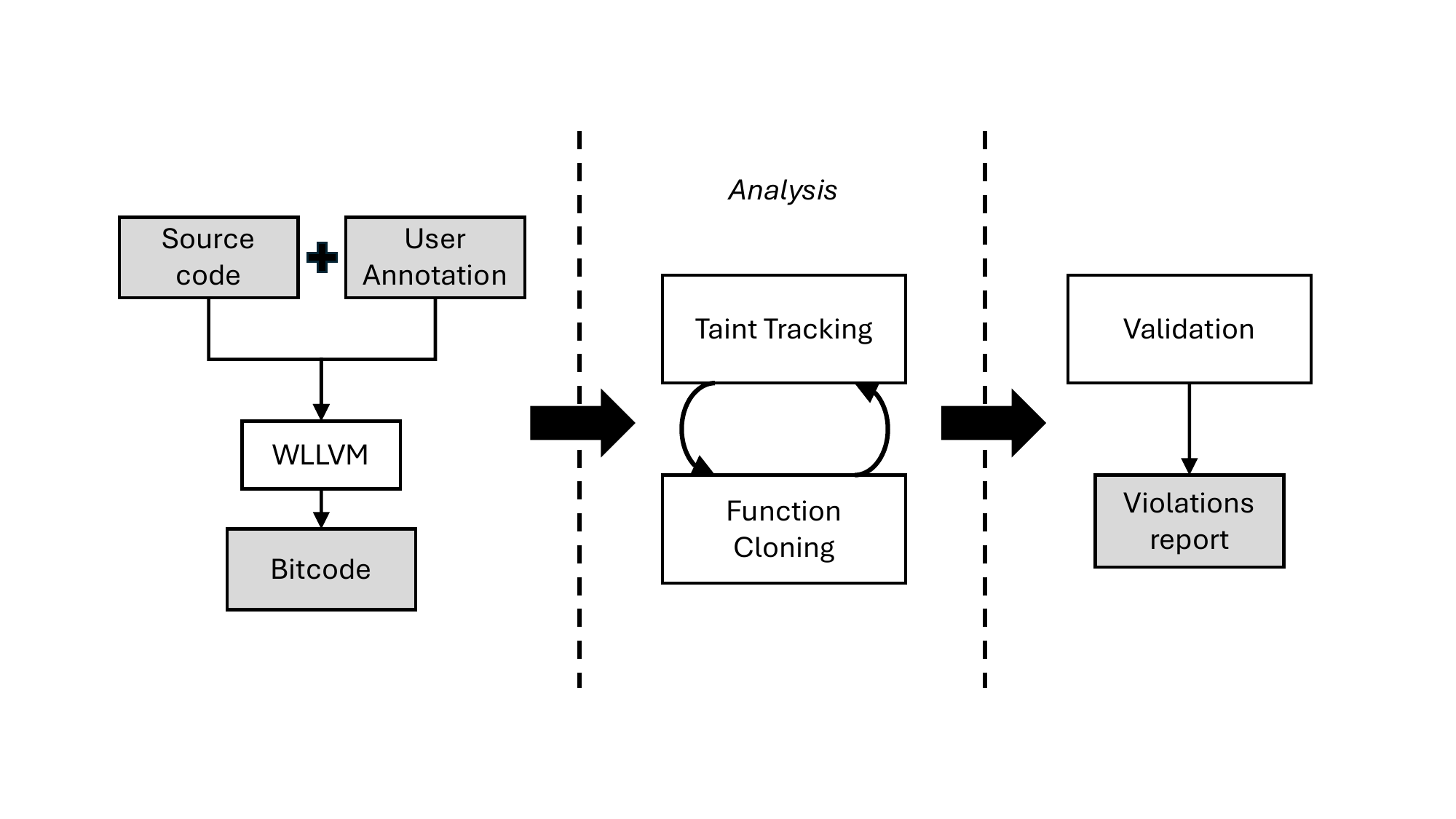}
    \caption{The \gls{BliMe} Linter Overview.}
    \label{fig:overview}
\end{figure}

The high-level design of \design is shown in \Cref{fig:overview}. The input to \design is the source code, annotated to mark  which data is sensitive, e.g., a secret key. Internally, \design{} is divided into two steps (\textsf{Analysis} and \textsf{Validation}). The \textsf{Analysis} step uses the sensitive data annotated by the developer as \emph{taint sources} and performs static taint-tracking analysis to identify where the taint can propagate throughout the program. The \textsf{Validation} step then identifies all locations in the program that might use tainted data in a non-BliMe-compliant manner, and creates a violations report. Note that since the analysis is not complete, the presence of violations in the final report does not necessarily mean that the executable will not run on BliMe; the violations might be a result of over-approximations of the analysis as discussed in \Cref{sec:overapprox}. On the other hand, if the violations report is empty, the executable is guaranteed to run on BliMe.

In the following subsections, we detail each part of our design.

\subsection{Developer Interface}
\label{sec:devIntf}

\begin{listing}
\begin{minted}
[
frame=lines,
framesep=3\fboxsep,
xleftmargin=18pt,
baselinestretch=1.1,
fontsize=\footnotesize,
linenos
]
{C}
__attribute__((blinded)) char x = 123;
__attribute__((blinded)) char *y = new char[10];
\end{minted}
\caption{Examples of using the \texttt{blinded} attribute.}\label{lst:annotations}
\end{listing}

Since \design relies on developer annotations to identify sensitive data, we introduce a new C++ Clang attribute (\texttt{blinded}) that the developer can attach to variables in the source code. 
The attribute works similar to the C/C++ type qualifiers such as \texttt{const} and can similarly be applied to primitive types and complex types such as structures and pointers.
The example in \Cref{lst:annotations}, shows a blinded primtive type (\texttt{char x}) and a pointer to an array of blinded characters.

We have extended the Clang front-end to recognize the \texttt{blinded} attribute and pass them on in the lowered LLVM \gls{IR}, which is then passed to the \textsf{Analysis} step.
As the LLVM compiler pipeline may drop unknown attributes, we schedule our analysis pass before other LLVM \gls{IR} passes that may discard them.

\begin{paper}
\end{paper}

\subsection{Whole-program LLVM}
When compiling C++ projects with more than a single \texttt{.cpp} file, build systems usually compile each file to a separate object file and then links all the object files together at the end to produce an executable. This means that analyses or optimizations \emph{across} modules at the \emph{LLVM IR} level are not possible. To solve this issue, we use \gls{wllvm}~\cite{wllvm}. During compilation, \gls{wllvm} generates LLVM bitcode for each compilation module and adds it to a custom ELF section in the corresponding object file. When the final executable is linked, \gls{wllvm} concatenates all bitcode sections from the linked files and adds them to a section in the ELF executable. A \gls{wllvm} tool can then be used to extract and link the bitcode. The result is a bitcode file that represents the entire program. This allows us to easily analyze the entire program (i.e., across modules) using a single input bitcode file.

\subsection{\textsf{Analysis}}
As shown in \Cref{fig:overview}, the \textsf{Analysis} step consists of two parts that are also performed iteratively: \textsf{Taint Tracking} and \textsf{Function Cloning}. \textsf{Taint Tracking} uses static value-flow tracking to propagate taint from the taint sources to the rest of the program. If there are any functions with newly tainted arguments, they are cloned, and the compiler then start another round of \textsf{Taint Tracking}. If there are no functions with newly tainted arguments, the compiler exits the Analysis step.

\subsubsection{\textsf{Taint Tracking}}\label{sec:taint-tracking}
We adapt the state-of-the-art SVF~\cite{SVF} tool to perform static value-flow tracking in the \textsf{Taint Tracking} step. From SVF, we first obtain a \gls{SVFG}, which shows the dependencies between values. %
We then traverse the \gls{SVFG} breadth-first and propagate taint from the taint sources to the rest of the program along the edges of the \gls{SVFG}.

\paragraph{Implicit Flows}
There is one case, however, which is not covered by SVF: \emph{implicit flows}, which are information flows from a condition to value assignments that depend on this condition. %
As a partial remedy for this gap, we propagate taint for LLVM \texttt{select} instructions. We use LLVM's internal def-use chains to detect such cases and propagate taint from the condition to the output.

\subsubsection{\textsf{Function Cloning}}\label{sec:function-cloning}
We introduce function cloning to improve the context-sensitivity of the analysis. A single function can be called from many call sites in the code. When taint propagates to the arguments of this function, we must track the taint within the function and, if the return value is tainted, propagate the taint back to the call site. However, doing this without function cloning can cause two issues:
\begin{enumerate}
    \item \textbf{Return value over-tainting:} If a function returns a tainted value at only a subset of its call sites, tainted return values will incorrectly propagate taint back to all call sites, even those not within this subset. For example, if a simple \texttt{add} function
    is called with tainted arguments at site \textcircled{a} and untainted arguments at site \textcircled{b}, taint will propagate back to the return value at both call sites. With function cloning, the function called at site \textcircled{a} will be a taint-propagating clone of the one called at site \textcircled{b}. Therefore, only the return value at site \textcircled{a} would become tainted, as expected. Another example is when two calls use tainted arguments, but only one should have a tainted return value.
    \begin{paper}
    \item \textbf{Forward call over-tainting:} 
    The situation described above can also occur for forward calls. A function called with tainted arguments can propagate these arguments to other functions calls within it. Consider the \texttt{add} example above. Without function cloning, if the \texttt{add} function calls another function, e.g., \texttt{copy}, with \texttt{add}'s tainted arguments, every call to \texttt{copy} will result in taint propagation, significantly increasing unnecessary taint.
    \end{paper}
\end{enumerate}

One drawback to function cloning is the increased code size caused by the additional functions. However, this is a reasonable trade-off considering the issues discussed above.

%% file: sections/5_eval.tex
\section{Evaluation}
\label{sec:eval}

To evaluate the soundness of \design{}, we first analyze the soundess of SVF. While the authors~\cite{SVF} do not claim any soundness guarantees, we believe it is safe to assume that SVF is sound with respect to \emph{explicit value flows} as per its design. %
Therefore, up until the first time tainted data flows into a condition (e.g., the condition in an if-then-else, or the condition in a select instruction), our analysis is sound. Beyond this point, we can no longer provide soundness guarantees. In other words, we make no claims that if violations are reported, that they are the \emph{only} violations. However, once a developer transforms a violating condition to a BliMe-compliant form, and reruns the analysis, our soundness guarantees will hold further along the program's execution paths. Through an iterative process, we can therefore claim soundness as follows: if \design{} does not report any violations, then the program is guaranteed to work on BliMe.

\begin{table}[ht]
\centering
\resizebox{\linewidth}{!}{%
\begin{tabular}{@{}lcccc@{}}
\toprule
\multicolumn{1}{c}{}        & \multicolumn{2}{c}{Analysis Results} & \multicolumn{2}{c}{Hardware Results} \\ \cmidrule(l){2-5} 
\multicolumn{1}{c}{Program} & Memory            & Branch            & Memory            & Branch           \\ \midrule
binary\_search & 1     & 3      & 0 & 2  \\
dijkstra       & 6     & 5      & 6 & 0  \\
dnn            & 0     & 0      & 0 & 0  \\
find\_max      & 0     & 1      & 0 & 1  \\
int\_sort      & 0     & 1      & 0 & 1  \\
kmeans         & 0     & 2      & 0 & 2  \\
matrix\_mult   & 0     & 0      & 0 & 0  \\
page\_rank     & 6     & 0      & 2 & 0  \\
PQ             & 3     & 7      & 0 & 3  \\ \midrule
label\_image   & 4,047 & 13,101 & 1 & 27 \\ \bottomrule
\end{tabular}%
}
\vspace{1em}
\caption{Results of running \design{} on the \gls{OISA} benchmarks. We compare the reported violations with results obtained dynamically on BliMe-Spike.}
\label{tab:results}
\end{table}

We further evaluate \design{} empirically against a debug-enabled implementation of \gls{BliMe} on the RISC-V Spike emulator, which we call BliMe-Spike. This debug-enabled implementation does not fault when it identifies BliMe violations, but instead produces warnings that identify the offending instructions. This allowed us to run each executable only once to obtain several violating instructions instead of having to ``fix'' and rerun the executable for every violation. 
However, due to its dynamic nature, BliMe-Spike has the following properties.
\begin{itemize}
    \item It can only identify violations that are on the execution path. This results in a list of violations that is a subset of that produced by \design{}.
    \item It intentionally does not propagate taint through implicit flows. It will correctly report a violation on a branching condition that depends on a blinded value, but will not propagate this violation to assignments performed within the violating then/else branches that would be unreachable with BliMe enforcements enabled. This is because BliMe-Spike does not have a high-level understanding of program semantics and therefore cannot reason about where the branches end. The other design choice would have been to propagate taint to \emph{all} assignments after the branching violation, but this would produce many false positives and would reduce the overall usefulness of the output.
    \item The same violation can result in many duplicate outputs. For example, a branching violation within a for loop that performs 100 iterations will produce 100 warnings instead of one. BliMe-Spike's output therefore requires post-processing to remove duplicates.
\end{itemize}

We selected two applications for the empirical evaluation: the benchmarks from \gls{OISA}~\cite{yu19}, and TensorFlow Lite~\cite{tensorflow2015-whitepaper}.

\subsection{OISA}
Yu et al. use a suite of benchmarks to evaluate \gls{OISA} by comparing the performance before and after the benchmarks are made constant-time. We therefore use the non-constant-time versions as a starting point for our evaluation. We compile the \gls{OISA} benchmarks, using the same sensitive inputs as Yu et al and annotating them with the \texttt{blinded} attribute to mark them as taint sources. We then compare our analysis results with the violations reported by BliMe-Spike. 
The results are shown in \Cref{tab:results}. We present two case studies from the OISA benchmarks that show the effectiveness of \design{}.

\subsubsection{\texttt{find\_max}}\label{sec:case-studies}
Our first case study uses the \texttt{find\_max} benchmark, which is a simple program that scans an array of secret values and finds the maximum value in that array. We show the source code for \texttt{find\_max} in \Cref{lst:findmax}.

\begin{listing}[ht]
\begin{minted}[
frame=lines,
framesep=2mm,
xleftmargin=18pt,
baselinestretch=1.1,
fontsize=\footnotesize,
linenos
]
{C}
void FindMax(
    __attribute__((blinded)) int arr[],
    int* max_idx,
    int* max_val){
  *max_val = -1;
  for (int i = 0; i < N; i++){
    if (arr[i] > *max_val){
      *max_idx = i;
      *max_val = arr[i];
    }
  }
}
\end{minted}
\caption{Source code of OISA benchmark \texttt{find\_max} for finding the maximum value.}\label{lst:findmax}
\end{listing}

As can be seen in the listing, we add the \texttt{blinded} attribute to the \texttt{arr} pointer parameter. This marks the data \emph{within} the array as blinded. Our analysis correctly identifies a BliMe violation on line 7. The violation is due to the use of a blinded value (\texttt{arr[i]}) in the condition of a branch. \Design{} has an optional feature to output LLVM \gls{IR} that contains taint attached to instructions as debugging metadata. The \gls{IR} output from the \texttt{find\_max} analysis is shown in \Cref{lst:findmax-IR}.

\begin{listing}[ht]
\begin{minted}[
frame=lines,
framesep=2mm,
xleftmargin=18pt,
baselinestretch=1.1,
fontsize=\footnotesize,
linenos
]
{llvm}
        i32* %
br i1 %
\end{minted}
\caption{LLVM \Gls{IR} of BliMe violation in the \texttt{find\_max} benchmark.}\label{lst:findmax-IR}
\end{listing}

Instructions marked with a \texttt{!t} represent blinded data. \texttt{\%8} is the pointer to \texttt{a[i]}, i.e., \texttt{\&a[i]}. Note that it is correctly not marked as blinded. The following load fetches the blinded data from the array and stores it in \texttt{\%9}, which is correctly marked as blinded. Taint is propagated by the analysis to \texttt{\%10}, which is then used as a condition in a branching instruction on the following line, resulting in a BliMe violation.

\subsubsection{\texttt{page\_rank}}
For the \texttt{page\_rank} benchmark, we examine the violation shown in \Cref{lst:pagerank}. \texttt{graph} is a pointer to a blinded struct representing the secret graph of web pages. This causes two violations on line 3. The first is when \texttt{numOutEdges} is loaded, and the second is when it stored (after incrementing). The exact taint propagation is shown more clearly in the LLVM \gls{IR} in \Cref{lst:pagerank-IR}. \texttt{\%10} loads the value of \texttt{e->src}, which is blinded. A pointer (\texttt{\%12}) is derived from this value and used to load \texttt{numOutEdges} (\texttt{\%13}). This is marked as a violation due to a load using a blinded address. The same pointer is then used again to store the incremented value on the line 7, resulting in another violation, this time due to a store using a blinded address.

\begin{listing}[ht]
\begin{minted}[
frame=lines,
framesep=2mm,
xleftmargin=18pt,
baselinestretch=1.1,
fontsize=\footnotesize,
linenos
]
{C}
for(int i = 0; i < numEdges; i++){
  Edge* e = &graph->edges[i];
  graph->vertices[e->src].numOutEdges++;
}
\end{minted}
\caption{Source code snippet of load/store violation in OISA benchmark \texttt{page\_rank}. \texttt{graph} is a pointer to a blinded struct. There are two violations reported on line 3: a store and a load.}\label{lst:pagerank}
\end{listing}

\begin{listing}[ht]
\begin{minted}[
frame=lines,
framesep=2mm,
xleftmargin=18pt,
baselinestretch=1.1,
fontsize=\footnotesize,
linenos
]
{llvm}
store i32 %
\end{minted}
\caption{LLVM \Gls{IR} of BliMe violations in the \texttt{page\_rank} benchmark.}\label{lst:pagerank-IR}
\end{listing}

\subsection{TensorFlow Lite}

For TensorFlow Lite, we used the image classification example, \texttt{label\_image}, marking the input image as the sensitive data. With BliMe-Spike, we obtained 27 branching violations and 1 memory access violation. 17 of the 27 branching violations were due to uses of the \texttt{std::max} and \texttt{std::min} function in libc. 17 of all violations were in the \texttt{gemmlowp} matrix multiplication library used internally by TensorFlow.

With \design{}, we faced challenges when running our Analysis iterations on the image classification example. Attempting to run the Analysis until no further iterations were required (i.e., function cloning produced no changes) resulted in a run-time greater than 70 hours and memory usage greater than 110GB. We discuss this in more detail in \Cref{sec:discussion:svf}. Our analysis was eventually killed due to lack of resources and we therefore decided to disable function cloning and rerun the analysis. Our memory usage was still high (\~80GB) but run-time was significantly lower (\~24 hours). As a result, however, the analysis was imprecise and there was a significant amount of taint over-approximation. This is evident in the results shown in \Cref{tab:results}. We discuss the challenges of using our approach for large programs in \Cref{sec:discussion}.

%% file: sections/6_discussion.tex
\section{Discussion \& Future Work}\label{sec:discussion}

\subsection{SVF execution}\label{sec:discussion:svf}
One challenge we faced when applying our solution to a large executable such as TensorFlow was SVF's very long run-time and high memory overhead, which align with reports on SVF's GitHub page. The cause of this is two-fold. The first is that pointer analysis takes a long time to cover the entire executable. The second is the \gls{SVFG} generation process itself.

One of the reasons the analysis takes a long time is that SVF builds an SVFG of the \emph{entire} program without considering where the taint sources are located. One possible solution is to limit the analysis to the parts of the program that might contain taint and build the SVFG in tandem with taint propagation.

\subsection{Taint over-approximation}\label{sec:overapprox}
As with any static analysis technique, our analysis must make conservative approximation to be computationally feasible (\Cref{sec:blime-vs-compiler}). We choose to use a conservative approach that over-approximates rather than under-approximates taint. This aligns with our goal of guaranteeing that the program will run if no violations are reported.

While this is sufficient to guarantee BliMe compatibility for a subset of programs, it may result in false negatives. 
We implemented function cloning to achieve partial context sensitivity (\Cref{sec:function-cloning}) and eliminate some false negatives but leave further exploration of more accurate analyses as future work.
Additional source code annotations could also be added to allow programmers to explicitly denote data as non-blinded to limit the complexity of the analysis.

\subsection{Applicability}
Although we focus in this paper on BliMe, \design{} is applicable to any system that requires data-oblivious execution. One prominent example is \gls{OISA}~\cite{yu19}, which enforces a similar policy to BliMe. Furthermore, BliMe linter's analysis is useful even for software that runs on ordinary processors without BliMe or other data-oblivious execution extensions because it can hep developers reason about how their software handles sensitive data and identify potential leaks.

\subsection{Transformations}\label{sec:transformations}
\Design provides developers with valuable information about potential side channel vulnerabilities in their programs. Developers can then use this information to manually transform the programs to remove these vulnerabilities. However, with a large volume of vulnerabilities, manual transformations can be cumbersome and time-consuming. Therefore, a natural next step for future work is to extend the compiler to perform these transformations automatically. While this might not be feasible in all cases (e.g., if static analysis cannot identify proper bounds for a blinded pointer), it can significantly reduce the amount of manual labor required, especially if a program contains many easily-transformed violations such as conditional select statements.

One way to integrate transformations into \design{} is by using Constantine~\cite{borrelloConstantine2021}.
Constantine is a set of LLVM compiler extensions that perform \gls{CFL} and \gls{DFL} transformations to produce constant-time code. \gls{CFL} is done using a form of predicated execution, which computes a predicate for each branch, executes both branches, and masks the results using the predicate to obtain the correct result and discard the result from the incorrect branch. \gls{DFL} linearizes array accesses that use a tainted index by iterating over the entire array using a stride equal to the cache line size, and selecting only the value at the correct index using the predicate.

While Constantine's transformations produce constant-time code, we identify two challenges with its application to BliMe and \gls{OISA}:

\subsubsection{Array access expansion}\label{sec:array-access}
Constantine's array access expansion results in binaries that are only constant-time on \glspl{CPU} with a cache line size matching the chosen stride. Furthermore, the offset of each access from the beginning of the cache line is derived from the sensitive index. On \gls{BliMe} and \gls{OISA} hardware, this will cause all the resulting memory access addresses to also be tainted and thus cause the instructions to fault. We can solve this by modifying the array access expansion transform to use a stride of one, which results in no sensitive offset being added to each address and the addresses remain untainted.

\subsubsection{Select transform}
Constantine uses x86-specific optimizations such as the \texttt{cmov} instruction. The RISC-V \gls{ISA} does not have such an instruction, which causes \gls{IR} \texttt{select} instructions to be lowered to conditional branching instructions in assembly. This means that integration with Constantine must include special handling for \gls{IR} select instructions.

%% file: sections/7_related.tex
\section{Related Work}

As mentioned in \Cref{sec:transformations}, Constantine~\cite{borrelloConstantine2021} is a state-of-the-art constant-time compiler. While Constantine's transformations are complementary to \design{}, Constantine also uses taint tracking to identify potential violations. However, it uses \emph{dynamic} analysis to propagate taint, which involves running the target program several times with different sets of realistic inputs and tracking where the taint propagates throughout each execution. The downside to this approach is that, contrary to \design{}, Constantine's analysis is not sound as it only considers the observed execution paths and misses information flows that only occur on paths that were not exercised.

%% file: sections/8_conclusion.tex
\section{Conclusion}

Manually identifying side channels in programs is difficult. \Design{} facilitates this process by performing static taint-tracking analysis and reporting potential violations. This allows developers to focus only on problematic areas in their software. Through an iterative process, we show that \design{} can help developers eliminate all non-compliant code. However, as with most static analysis techniques, there is room to improve precision. In addition, automatic code transformations are a promising line of future work.